\documentclass[10pt]{iopart}
\usepackage{iopams,bm,cite,color,graphicx,braket,mathptmx}
\eqnobysec
\newcommand{\openone}{\leavevmode\hbox{\small1\normalsize\kern-.33em1}}

\begin{document}

\title{The Wigner flow on the sphere}
\author{Popo~Yang$^{1}$, Iv\'an~F~Valtierra$^{2}$,
  Andrei~B~Klimov$^{2}$, Shin-Tza~Wu$^{3}$,
  Ray-Kuang~Lee$^{1}$, Luis~L~S\'anchez-Soto$^{4,5}$ and
Gerd~Leuchs$^{5,6,7}$}

\begin{abstract}
  We derive a continuity equation for the evolution of the SU(2)
  Wigner function under nonlinear Kerr evolution. We give explicit
  expressions for the resulting quantum Wigner current, and discuss
  the appearance of the classical limit. We show that the global
  structure of the quantum current significantly differs from the
  classical one, which is clearly reflected in the form of the
  corresponding stagnation lines.
\end{abstract}

\address{$^{1}$Department of Physics, National Tsing Hua University,
  Hsinchu 300, Taiwan}

\address{$^{2}$Departamento de F\'{\i}sica, Universidad de
  Guadalajara, 44420~Guadalajara, Jalisco, Mexico}

\address{$^{3}$Department of Physics, National Chung Cheng
  University, Chiayi 621, Taiwan}

\address{$^{4}$Departamento de \'Optica, Facultad de F\'{\i}sica,
  Universidad Complutense, 28040~Madrid, Spain}

\address{$^{5}$ Max-Planck-Institut f\"ur die Physik des Lichts,
  Staudtstra{\ss}e 2, 91058~Erlangen, Germany}

\address{$^{6}$ Department of Physics, 
University of Erlangen-Nuremberg, Staudtstra\ss e 7 B2, 
91058 Erlangen, Germany}

\address{$^{7}$ Physics Department, Centre for Research in Photonics, 
University of Ottawa, Advanced Research Complex, 25 Templeton, 
Ottawa ON Canada, K1N 6N5}


\section{Introduction}

In classical statistical mechanics, an ensemble of particles is
described by a distribution function $f(x,p|t)$ that depends on the
phase-space variables $x$ and $p$ and evolves in time. The
corresponding dynamics is governed by the Liouville
equation~\cite{Arnold:1989aa}, which asserts that for conservative
forces $f(x,p|t)$ is constant along the trajectories of the system. In
other words, the local density of points traveling through phase-space
is constant with time. This conservation can be succinctly summarized
as a continuity equation
\begin{equation}
  \frac{\partial f(x,p|t)}{\partial t} =-
  \nabla \cdot \mathbf{J}(x,p|t) \, ,
  \label{eq:Liouflow}
\end{equation}
where $\mathbf{J}(x,p|t)$ is a probability current. Indeed, this flow
is regular~\cite{Berry:1978aa} and largely determined by location and
nature of its stagnation points; i.e, those points for which
$\mathbf{J}=0$.  For conservative systems, the form of $J(x,p|t)$
immediately follows from the corresponding Poisson brackets.

This scenario can be extended to more general systems admitting a
dynamical symmetry group. This enables the construction of a phase
space $\mathcal{M}$ as an appropriate homogeneous
manifold~\cite{Stratonovich:1956qc,Berezin:1975mw}. This classical
formulation associates a probability with every point
$\Omega \in \mathcal{M}$. However, in the quantum domain the
uncertainty principle does not allow one to attribute a state to a
single point in phase space~\cite{Hillery:1984aa,Lee:1995aa,
  Schroek:1996aa,Schleich:2001aa,QMPS:2005aa}. Because of this
fundamental difference, there is no unique way of defining a quantum
probability distribution. The Wigner function $W_{\varrho}(\Omega )$
is perhaps the closest to the analogous counterpart. Note 
that, although the Wigner function has the correct marginal
probability distributions, it can itself be negative.

In the phase-space approach, every observable $\hat{A}$ is mapped onto
a function $W_{A} (\Omega)$ (called its Weyl symbol). In particular,
the Weyl symbol of the density matrix is precisely the Wigner function
and its time evolution reads
\begin{equation}
  \partial_{t}W_{\varrho}(\Omega |t ) =
  \{  W_{\varrho}(\Omega ),W_{H}(\Omega ) \}_{M} \, ,
  \label{ee1}
\end{equation}
where $W_{H}(\Omega)$ is the symbol of the Hamiltonian and the Moyal
bracket $\{ \cdot, \cdot \}_{M}$ is the image of the quantum
commutator [times $(\rmi \hbar)^{-1}$] under the Weyl
map~\cite{Moyal:1949aa}.  The resulting partial differential equation
contains, in general, higher-order derivatives, which significantly
complicate the search for an exact solution. However, it admits a
natural expansion in powers of a semiclassical parameter
$\varepsilon \ll 1$ that characterizes the strength of quantum
fluctuations in the system.  This parameter depends on the dynamical
symmetry and, roughly speaking, is the inverse of the number of
excitations~\cite{Klimov:2017aa}. To the lowest order in
$\varepsilon$, equation (\ref{ee1}) is of the Liouvillian form
\begin{equation}
  \partial_{t}W_{\varrho}(\Omega ) =
  \varepsilon \{W_{\varrho}(\Omega ),W_{H}(\Omega )\} +
  O(\varepsilon^{3}) \, ,
  \label{ee2}
\end{equation}
where now $\{ \cdot , \cdot \} $ is the Poisson bracket in the
manifold $\mathcal{M}$.  The semiclassical or truncated Wigner
approximation (TWA)~\cite{Ozorio:1998aa,Klimov:2002cr,
  Dittrich:2010aa,Polkovnikov:2010aa} consists in disregarding the
higher order terms, so that $W_{\varrho}(\Omega |t) \simeq
W_{\varrho}(\Omega(-t)|0)$, where $\Omega (t)$ are classical
trajectories generated by $W_{H}(\Omega )$.

It has been pointed out~\cite{Bauke:2011aa,Steuernagel:2013aa,
  Kakofengitis:2017aa}, that one can construct a Wigner
current~\footnote{We will mainly call the quantity $\mathbf{J}$ the
  Wigner current. However, it can also be interpreted as
  quasiprobability flow. For this reason, the designation Wigner flow
  has been used in the literature before. Note though that, as
  discussed in \cite{Oliva:2018aa}, no flow (in the sense of mapping
  of a distribution along trajectories) exists in the quantum domain.}
in such a way that the evolution can be mapped as a continuity
equation very much analogous to (\ref{eq:Liouflow}). Surprisingly,
this Wigner current, which is the equivalent of the classical
Liouville flow, has, so far, not been studied in great
detail~\cite{Donoso:2001aa, Hughes:2007aa,
  Kakofengitis:2017ab,Friedman:2017aa}.  The form of the current, and
especially the behavior in the vicinity of its stagnation points, can
be used for the characterization of the quantumness of the evolution
(see also references~\cite{Skodje:1989aa,Veronez:2013aa,
  Veronez:2016aa}, where the stagnation points of the Husimi $Q$
function were studied).

In this paper, we extend these ideas to spinlike systems, where the
classical phase space is the unit sphere. We stress that this is not a
mere academic curiosity, since the underlying SU(2) symmetry plays a
pivotal role in numerous models in physics~\cite{Chaturvedi:2006vn}.

In the spirit of equations (\ref{ee1}) and (\ref{ee2}), we introduce
in a natural way the classical and quantum Wigner currents. We will
show, using the simplest example of nonlinear Kerr dynamics, that the
global structure of the quantum Wigner current significantly differs
from the classical one.  Such a difference is clearly observable even
during the short-time evolution of semiclassical states (specified by
localized distributions in phase space), when the Wigner distribution
can still be well described in terms of the semiclassical
approximation. In other words, the Wigner current allows us to
distinguish between quantum and semiclassical dynamics, while the
distributions evolved according to the Moyal and Poisson brackets are
still quite similar. The stagnation points/lines of the classical
Wigner current are basically determined by the zeros of the
semiclassicaly evolved Wigner distribution. Therefore, the structure
of the stagnation lines can be used for the analysis of the
quantumness of the phase-space dynamics in the semiclassical limit.
Furthermore, an extra benefit of bringing the Wigner current into play
is that it can give a compelling visual representation of how
nonclassical features arise during the evolution.

\section{Wigner function on the sphere}

We consider a system whose dynamical symmetry group is SU(2). As
heralded in the Introduction, we follow the ideas in
references~\cite{Stratonovich:1956qc,Berezin:1975mw} to work out
quasiprobability distributions on the sphere satisfying all the
pertinent requirements. This construction was generalized by
others~\cite{Agarwal:1981bd,Brif:1998if,Heiss:2000kc,
  Klimov:2000zv,Klimov:2008yb} and has proved to be very useful in
visualizing properties of spinlike systems~\cite{Dowling:1994sw,
  Atakishiyev:1998pr,Chumakov:1999sj,Chumakov:2000le}.

The corresponding Lie algebra $\mathfrak{su} (2)$ is spanned by the
operators $\{ \hat{S}_{x}, \hat{S}_{y}, \hat{S}_{z} \}$ satisfying the
angular momentum commutation relations
\begin{equation}
  [\hat{S}_{x}, \hat{S}_{y} ] = \rmi \hat{S}_{z} \,,
\end{equation}
  and cyclic permutations (in units $\hbar =1$, which will be used
  throughout). The Casimir operator is
  $\hat{\mathbf{S}}^{2} = \hat{S}_{x}^{2} + \hat{S}_{y}^{2}+
  \hat{S}_{z}^{2} = S (S+1) \openone$, so the eigenvalue $S$ (which is
  a nonnegative integer or half integer) labels the irreducible
  representations (irreps). We take a fixed irrep of spin $S$, with a
  $2S+1$-dimensional carrier space $\mathcal{H}_{S}$ spanned by the
  standard angular momentum basis
  $\{ |S m\rangle , m= -S, \ldots, S\}$, whose elements are
  simultaneous eigenstates of $\hat{\mathbf{S}}^{2}$ and
  $ \hat{S}_{z}$:
\begin{equation}
  \hat{\mathbf{S}}^{2} |S, m \rangle = S (S+1) |S, m \rangle \, ,
  \qquad
  \hat{S}_{z} |S, m\rangle = m |S, m \rangle \, .
\end{equation}

The highest weight state is $|S,S\rangle$ and it is annihilated by the
ladder operator $\hat{S}_{+}$ (with
$\hat{S}_{\pm} = \hat{S}_{x} \pm \rmi \hat{S}_{y}$). The isotropy
subgroup (i.e., the largest subgroup that leaves the highest weight
state invariant) consists of all the elements of the form
$\exp( i \chi \hat{S}_{z})$, so it is isomorphic to U(1). The coset
space is then SU(2)$/$U(1), which is just the unit sphere
$\mathcal{S}_{2}$ and it is the classical phase space, the natural
arena to describe the dynamics.

The SU(2) coherent states $|\Omega \rangle $ (with
$\Omega =(\theta ,\phi )\in \mathcal{S}_{2}$) are defined,
up to a global phase, by the action of the displacement
operator~\cite{Perelomov:1986ly}
\begin{equation}
  \hat{D}(\Omega )= \exp \left [
    \case{1}{2}\theta (\hat{S}_{+}\rme^{-\rmi\phi} -
    \hat{S}_{-}\rme^{\rmi\phi} ) \right ]
  \label{eq:cs1}
\end{equation}
on the highest weight state, with explicit expression in terms of $\Omega $
given by 
\begin{eqnarray}
|\Omega \rangle & = & \hat{D}(\Omega )|S,S\rangle  \nonumber \\
&=&\sum_{m=-S}^{S}\sqrt{\frac{(2S)!}{(S-m)!(S+m)!}}[\cos (\theta
/2)]^{S+m}[\sin (\theta /2)]^{S-m}\,\rme^{-\rmi m\phi}|S,m\rangle
    \,. \nonumber \\
\label{CS}
\end{eqnarray}

Operators acting in a $\mathcal{H}_{S}$ can be mapped onto functions
on $\mathcal{S}_{2}$ by means of the Stratonovich-Weyl kernel. It can
be concisely defined as~\cite{Varilly:1989ud}
\begin{equation}
  \hat{w}(\Omega ) = \sqrt{\frac{4\pi}{2S+1}}
  \sum_{K=0}^{2S} \sum_{q=-K}^{K}
  Y_{Kq}^{\ast}(\Omega ) \, \hat{T}_{Kq}^{S}\,,
  \label{kernel}
\end{equation} 
where $Y_{Kq}(\Omega )$ are the spherical harmonics, $\ast$ indicates
complex conjugation, and $\hat{T}_{Kq}^{S}$ are the irreducible tensor
operators~\cite{Fano:1959ly,Blum:1981rb}
\begin{equation}
  \hat{T}_{Kq}^{S}=\sqrt{\frac{2K+1}{2S+1}}
  \sum_{m,m^{\prime}=-S}^{S} C_{Sm,Kq}^{Sm^{\prime}}\,
  |S,m^{\prime}\rangle \langle S,m|\,,
\end{equation}
$C_{Sm,Kq}^{Sm^{\prime}}$ being the corresponding Clebsch-Gordan
coefficient~\cite{Varshalovich:1988ct}. The symbol $W_{A}$ of an
operator $\hat{A}$ is then defined as
\begin{equation}
  W_{A}(\Omega )=\Tr [ \hat{A} \, \hat{w} ( \Omega ) ]\,.  \label{Wff}
\end{equation}
Since the tensors $\hat{T}_{Kq}^{S}$ constitute an orthonormal basis
for the operators acting on $\mathcal{H}_{S}$, any observable
$\hat{A}$ can be expanded as
\begin{equation}
  \hat{A}=\sum_{K=0}^{2S} \sum_{q=-K}^{K}
  A_{Kq}\,\hat{T}_{Kq}^{S}\,,
  \label{A}
\end{equation}
with $A_{Kq}=\Tr [ \hat{A} \hat{T}_{Kq}^{S \dagger} ]$, $\dagger$
standing for Hermitian conjugation. Therefore, the symbol of $\hat{A}$
can be expressed as the sum of symbols of the tensor components
\begin{equation}
  W_{A}(\Omega )=\sqrt{\frac{4\pi}{2S+1}}
  \sum_{K=0}^{2S}\sum_{q=-K}^{K}A_{Kq} \,Y_{Kq}^{\ast}(\Omega ) \,.
  \label{eq:symb}
\end{equation}
As some relevant examples we shall need in what follows we quote 
\begin{equation}
\begin{array}{rcl}
  \label{eq:symbols}
  \hat{S}_{i} & \mapsto & W_{S_{i}}(\Omega )
 = \sqrt{S(S+1)} \;n_{i}, \\ 
  \{\hat{S}_{i},\hat{S}_{j}\} & \mapsto & W_{\{S_{i},S_{j}\}}(\Omega)
  =  C_{S}\,n_{i}n_{j}\,, \\ 
  \hat{S}_{i}^{2} & \mapsto & W_{S_{i}^{2}}(\Omega )
 =\frac{1}{2}C_{S}\left(
n_{i}^{2}-\frac{1}{3}\right) +\frac{1}{3}S(S+1),
\end{array}
\end{equation}
where the Latin indexes run the values $\{i,j\}\in x,y,z$,
$\mathbf{n}= (\sin \theta \cos \phi ,\sin \theta \sin \phi ,\cos
  \theta )^{t}$ is a unit vector in the direction of spherical angles
  $(\theta ,\phi ) \in \mathcal {S}_{2}$, and
$ C_{S}=[S(S+1)(2S-1)(2S+3)]^{1/2}$.

The Wigner function is the symbol of the density operator
$\hat{ \varrho}$. It is SU(2) covariant: under the action of a
$2S+1$-dimensional irrep of SU(2) given by the matrix
$\hat{R} (\Omega)$ [that is, $\hat{\varrho}^{\prime} =
\hat{R} (\Omega^{\prime}) \, \hat{\varrho} \,
\hat{R}^{-1} (\Omega^{\prime})$], $W_{\varrho}(\Omega)$
experiences the transformation
\begin{equation}
W_{\varrho^{\prime}} (\Omega) = W_{\varrho} (R^{-1} \Omega) \, ,
\end{equation}
so that it follows rotations rigidly without changing its form. In
addition, we have  the overlap relation
\begin{equation}
  \Tr (\hat{\varrho}\hat{A}) =
  \frac{2S+1}{4\pi}\int_{\mathcal{S}_{2}}\rmd\Omega
  \, W_{\varrho}(\Omega ) \, W_{A}(\Omega ) \,,
  \label{traslape}
\end{equation}
where $\rmd\Omega =\sin \theta \rmd\theta \rmd\phi $ is the invariant
measure in $\mathcal{S}_{2}$.

For a coherent state $|\Omega_{0}\rangle $, the Wigner function
can be computed directly from the definition (\ref{Wff}) by
taking into account  that 
\begin{equation}
  \langle \Omega_{0}|\hat{T}_{Kq}^{S}|\Omega_{0}\rangle =
  (2S)! \sqrt{\frac{4\pi}{(2S-K)!(2S+K+1)!}} \,
  Y_{Kq}(\Omega_{0}) \, .
\end{equation}
The final results thus reads
\begin{equation}
  W_{\Omega_{0}} (\Omega ) = (2S)! \sum_{K=0}^{2S}
  \sqrt{\frac{2S+1}{(2S-K)!(2S+K+1)!}}\, P_{K}(\cos \zeta ) \, ,
\end{equation}
where
$\cos \zeta = \cos \theta \cos \theta_{0} + \sin \theta \sin
\theta_{0}\cos (\phi -\phi_{0})$ and $P_{K}(\omega )$ are the Legendre
polynomials.

To conclude, we stress that this approach assume a fixed $S$. In some
instances, as in polarization optics, a superposition of different $S$
arise~\cite{Muller:2012ys,Muller:2016aa}.  The formalism can be
generalized to cover this more general situation~\cite{Tilma:2016aa}.

\section{Dynamics and Wigner current  on the sphere}

The exact evolution equation for the Wigner function
$W_{\varrho}(\Omega )$ has been obtained in \cite{Klimov:2002cr}. For
arbitrary Hamiltonians [living in a $(2S+1)$-dimensional
representation of the universal enveloping algebra of
$\mathfrak{su}(2)$] the expressions are quite involved. For
simplicity, in what follows, we restrict ourselves to two simple
examples of great interest in applications.

\subsection{Linear Hamiltonians}

First of all, we consider the dynamics generated by linear Hamiltonians 
\begin{equation}
\hat{H}_{L}=\sum_{i}a_{i}\hat{S}_{i}
\end{equation}
whose symbol can be directly inferred from (\ref{eq:symbols}). The
exact phase-space evolution is given by the first-order partial
differential equation
\begin{equation}
  \partial_{t}W_{\varrho}(\Omega |t)=\sum_{i}a_{i}
  \{W_{\varrho}(\Omega |t),n_{i}\} \, ,
  \label{Lin Ev}
\end{equation}
where
\begin{equation}
  \{f,g\}=\frac{1}{\sin \theta}
  \left( \partial_{\phi}f\,\partial_{\theta}g -
    \partial_{\theta}f\,\partial_{\phi}g\right)
\end{equation}
is the Poisson bracket on the sphere $\mathcal{S}_{2}$. The evolution
for the Wigner function is
\begin{equation}
  W_{\varrho}(\Omega |t)=W_{\varrho}(\Omega (-t)|0)\,,
  \label{Csol}
\end{equation}
where $\Omega (t)$ denotes classical trajectories, which are solutions
of the classical Hamiltonian equations. It thus corresponds to a
rotation of the initial distribution.

Next, we observe that if the evolution can be recast in the form $
\partial_{t}W_{\varrho}(\Omega |t)=\{A,B\}$, it can be interpreted as a
continuity equation with current given by 
\begin{equation}
  J_{\phi}=-A\partial_{\theta}B\, ,
  \qquad J_{\theta}=\frac{1}{\sin \theta}
  A \partial_{\phi}B \, .
  \label{Jgen}
\end{equation}
Accordingly, the linear dynamics is generated by 
\begin{eqnarray}
  J_{\theta} & = & \frac{1}{\sin \theta}
  W_{\varrho}(\Omega |t) \sum_{i}a_{i}\partial_{\phi}n_{i}\, ,
 \nonumber \\
& &  \label{flowG} \\
  J_{\phi} &=&-W_{\varrho}(\Omega |t)\sum_{i}a_{i}\,
 \partial_{\theta} n_{i}\,.  \nonumber
\end{eqnarray}
Since for these linear Hamiltonians the exact evolution is the
classical Liouville equation~\cite{Bayen:1978aa}, the quantum and
classical currents are just the same.

For the particular case of $\hat{H}_{L}=\omega \hat{S}_{z}$ the
resulting components of Wigner current are:
\begin{eqnarray}
  J_{\theta} & = &  0 \,,  \nonumber \\
             & &  \label{Jgen1} \\
  {J}_{\phi} & = & {\omega \sin \theta \;
                   W_{\varrho} (\theta ,\phi -\omega t|0) \, .}  \nonumber
\end{eqnarray}
In the supplemental material, we present an animation of this current
for an initial coherent state.

\subsection{Kerr dynamics}

For quadratic Hamiltonians, we content ourselves with the
simplest case of the so-called Kerr medium~\cite{Kitagawa:1993aa,
Agarwal:1997aa}, which is described by 
\begin{equation}
  \hat{H}=\chi \hat{S}_{z}^{2} \, .
  \label{HKerr}
\end{equation}
The ensuing dynamics has been examined in terms of the standard
position-momentum phase space~\cite{Corney:2008aa,Rigas:2013aa} and
the associated Wigner current has been recently
discussed~\cite{Oliva:2018ab}. For the SU(2) Wigner function the
evolution equation turns out to be~\cite{Klimov:2000zv,Klimov:2002cr}
\begin{equation}
  \partial_{t}W_{\varrho}(\Omega |t)= -
  \frac{\chi}{\epsilon} \cos \theta \
  \hat{\Gamma}(\theta ,\mathcal{L}^{2}) \,
  \partial_{\phi}W_{\varrho}(\Omega|t) \,,
  \label{KerrEE}
\end{equation}
where $\varepsilon= 1/(2S+1)$ and the operator
$\hat{\Gamma} (\theta ,\mathcal{L}^{2}) $ is
\begin{equation}
  \hat{\Gamma}(\theta ,\mathcal{L}^{2}) = \frac{1}{2}
  \Phi (\mathcal{L}^{2}) - \frac{\epsilon^{2}}{2}
  (1+2 \tan \theta \, \partial_{\theta}) \Phi^{-1} (\mathcal{L}^{2})
  \, .
  \label{G}
\end{equation}
Here, $\Phi (\mathcal{L}^{2})$ is
\begin{equation}
  \Phi (\mathcal{L}^{2}) = \left [ 2-\epsilon^{2}
    (2 \mathcal{L}^{2} + 1) +  2 \sqrt{1-\epsilon^{2}
      (2\mathcal{L}^{2}+1)+\epsilon^{4}\mathcal{L}^{4}}\right]^{1/2}
  \, ,
\end{equation}
and $\Phi^{-1} (\mathcal{L}^{2})$ its inverse. Both are functions
solely of $\mathcal{L}^{2}$, which is a differential realization of
the Casimir operato on $\mathcal{S}_{2}$
\begin{equation}
  \label{eq:Casi}
  \mathcal{L}^{2} = - \left (
    \partial_{\theta \theta}+
    \cot \theta \,\partial_{\theta}+
    \frac{1}{\sin^{2}\theta}\partial_{\phi \phi}\right) \, ,
\end{equation}
and, consequently, we have
$\mathcal{L}^{2} Y_{Kq}(\Omega )=K(K+1)Y_{Kq}(\Omega )$. Note also
that the term between parentheses in (\ref{eq:Casi}) is the
Laplace-Beltrami operator in $\mathcal{S}_{2}$.
\begin{figure}[t]
  \centering{\includegraphics[width=.98\columnwidth]{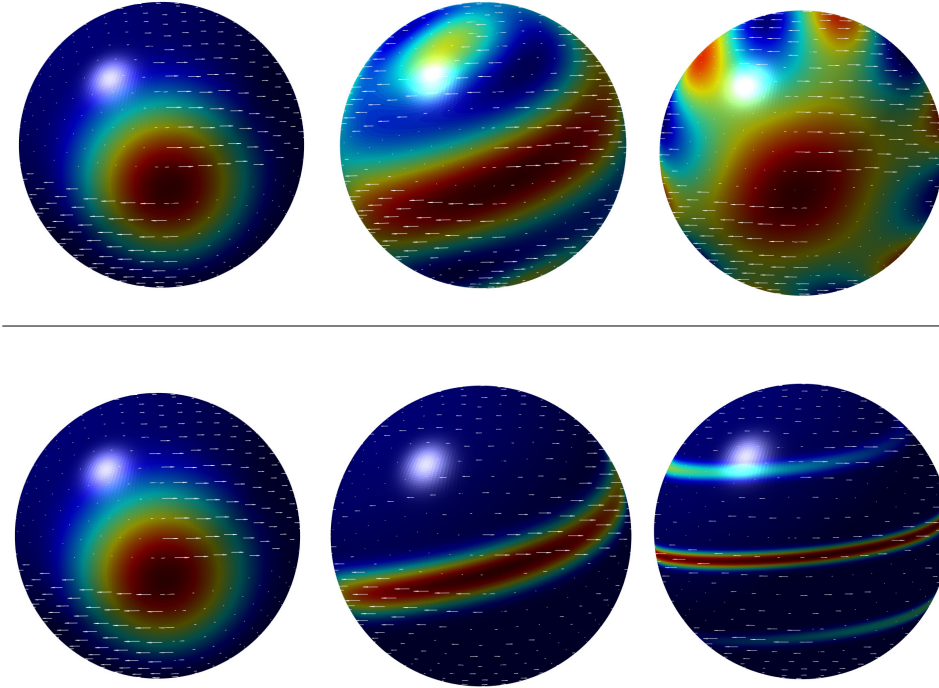}}
  \caption{Snapshots of Kerr dynamics for an initial atomic coherent
    state ($ S=10$) located in the equator at the dimensionless times
    $\tau =0$, $ \tau =0.32$ and $\protect\tau =1.5$. Upper panel,
    quantum dynamics; lower panel, semiclassical evolution.}
  \label{fig:flows}
\end{figure}

Equation~(\ref{KerrEE}) can be represented in terms of the Poisson
brackets as follows
\begin{equation}
  \partial_{t}W_{\varrho}(\Omega |t)=2\epsilon \chi \left\{ \hat{\Gamma}
    (\theta ,\mathcal{L}^{2})\,W_{\varrho}(\Omega ),\frac{1}{4\epsilon^{2}}
    \cos^{2}\theta \right\} \,.  \label{EEP1}
\end{equation}
Actually, the operator $\hat{\Gamma}(\theta ,\mathcal{L}^{2})$ is
responsible for the quantum deformation of the distribution. The
current can be immediately found from~(\ref{Jgen}):
\begin{eqnarray}
  J_{\phi}(t) &=&\chi \epsilon^{-1}\sin \theta \cos \theta \;\hat{\Gamma}
                  (\theta ,\mathcal{L}^{2})\;W_{\varrho}(\Omega |t)  \nonumber \\
              && \\
  J_{\theta}(t) &=&0\,.  \nonumber
\end{eqnarray}
The nonzero components of the current can be recast as 
\begin{equation}
J_{\phi}(t)=\sin \theta \hat{U}(t)\frac{1}{\sin \theta}J_{\phi}(t=0),
\end{equation}
where
\begin{equation}
  \hat{U}(t)=\exp \left[ -\frac{\chi t}{\epsilon}\cos \theta
    \,\hat{\Gamma} (\theta ,\mathcal{L}^{2}) \ \partial_{\phi}\right] 
\end{equation}
 is the evolution operator in phase space; that is,
$W_{\varrho}(\Omega |t)=\hat{U}(t)W_{\varrho}(\Omega |t=0)$.

\begin{figure}[tbp]
\centering{\includegraphics[width=\columnwidth]{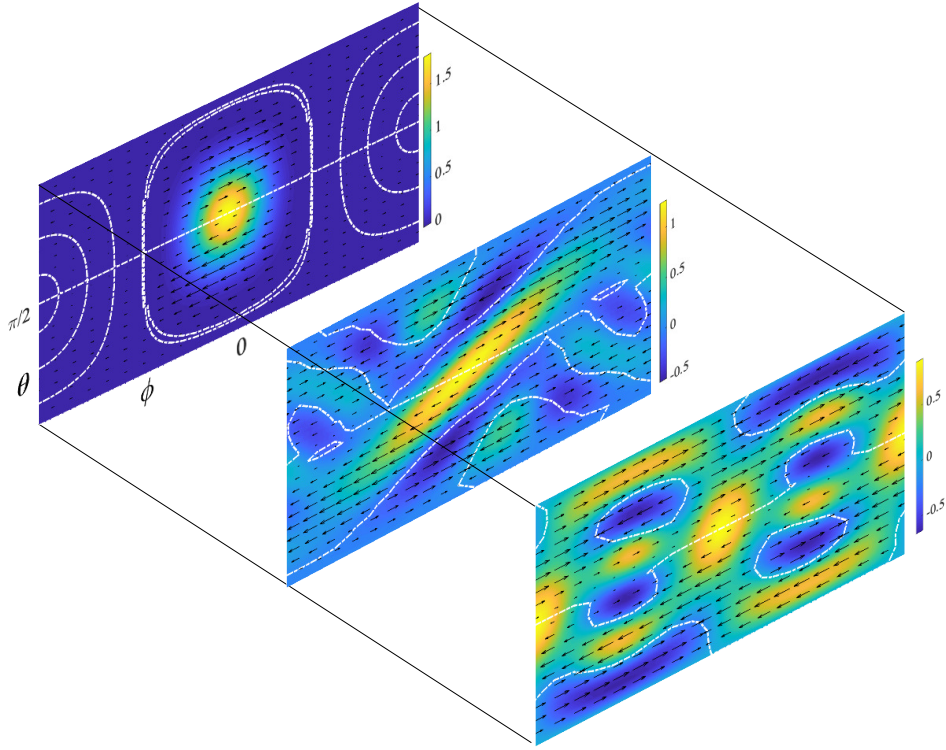}} .
\caption{The quantum current in figure~\ref{fig:flows}, but plotted in
  the plane (black arrows) for the same three times. The stagnation
  lines (white curves) separate regions of positive and negative
  values of the Wigner function.}
\label{fig:plane}
\end{figure}

In figure~\ref{fig:flows} we plot the quantum current for an initial
coherent state on the equator ($\theta =\pi /2, \phi =0$). We have
chosen three different dimensionless times $\tau =\chi t$
corresponding to $\tau =0$, $\tau =0.32$ (close to the best squeezing
time) and $\tau =1.5$ (close to the appearance of two-component
Schr\"{o}dinger cats) for the case $S=10$. The white arrows represent
the current $J_{\phi}$. At $\tau =0$ , the size and position of the
arrows clearly indicate the direction of the deformation of the Wigner
function: in the vicinity of the initial maximum, the \textit{laminar
  flow} with increasing speed towards polar regions leads to the
squeezing of the distribution along transverse directions for short
times $\tau \sim S^{-1/2}$. Such a deformation is actually reflected
in a real squeezing of $\hat{S}_{x}$ and $\hat{S}_{y}$ components. In
addition, first signs of the quantum interference are observed. For
cat times, $\tau \sim 1$ , the structure of the quantum current is
quite complicated: multiple regions where the current changes
direction can be easily noticed. In the supplemental material, the
reader can find an animation of this current for an initial coherent
state.

To better appreciate the stagnation lines (recall that $J_{\theta}=0$
identically), in figure~\ref{fig:plane} we plot the Wigner current of
figure~\ref{fig:flows}, but now in the plane. There is always a
trivial zero line at $\theta =\pi /2$. At the initial moment, the
stagnation lines separate regions of positive and negative values of
the Wigner function, as well as the minima of the negative
ripples. New zero lines around negative parts of the Wigner
distribution appear at the best squeezing time. Finally, nontrivial
stagnation lines take the form of closed curves rounding minima of the
interference pattern. These stagnation lines thus provide a
complementary picture of quantum interference in phase space.

\subsection{Semiclassical limit}

The semiclassical limit in spinlike systems is related to large value
of spin, naturally characterized by the parameter $\epsilon \ll
1$. The semiclassical states are usually associated with smooth and
localized (with extension of order $\sqrt{S}$) phase-space
distributions~\cite{Valtierra:2017aa}. Algebraically, the density
matrix of semiclassical states is decomposed only on low rank tensors
with $K\lesssim \sqrt{S}$ in equation~(\ref{A}). The typical
semiclassical states are the spin coherent states (\ref{CS}).

The operator (\ref{G}) in the semiclassical limit tends to $\hat{\Gamma}
(\theta ,\mathcal{L}^{2})=1+O(\epsilon^{2})$, and the evolution
equation (\ref{EEP1}) takes the form of the classical equation of motion
corresponding to the Hamiltonian (\ref{HKerr}); viz,
\begin{equation}
\partial_{t}W_{\varrho}(\Omega |t)\simeq -\frac{\chi}{\epsilon}\cos
\theta \partial_{\phi}W_{\varrho}(\Omega |t)=2\epsilon \{W_{\varrho
}(\Omega ),W_{H}(\Omega )\}\,,
\end{equation}
where the symbol of the Hamiltonian is 
\begin{equation}
W_{H}(\Omega )\simeq \frac{\chi}{4\epsilon^{2}}\cos^{2}\theta \,.
\end{equation}
The solution is defined by the classical trajectories according to
equation (\ref{Csol}). Nevertheless, in this case different points of
the initial distribution evolve with different velocities, so that the
classical motion leads to a semiclassical deformation of the initial
distribution,
\begin{equation}
  W_{\varrho}\left( \Omega |t\right) \simeq
  W_{\varrho}\left( \theta ,\phi - \frac{\chi t}{\epsilon}\cos \theta
    \Bigl |t=0\right) = W_{\varrho}(\Omega (-t)|0)  \, .
  \label{Wcl}
\end{equation}

The evolution distorts the initial distribution but cannot convert
positive regions of the Wigner function into negative regions (and
vice versa) as follows from the conservation of local Poincare\'e
invariants under the action of Poisson bracket.

Such a deformation represents, for instance,  squeezing and is
generated by the semiclassical current
\begin{eqnarray}
J_{\theta}^{\mathrm{scl}}(\Omega ) &=&0,  \nonumber \\
&&  \label{cl SL} \\
  J_{\phi}^{\mathrm{scl}} (\Omega ) &=& \frac{1}{2} \frac{\chi}{\epsilon}
\sin (2\theta) \,  W_{\varrho}(\Omega (-t)| 0)  \,.  \nonumber
\end{eqnarray}

In the lower panel of figure~\ref{fig:flows} we plot this
semiclassical current at the same times as for the quantum case. At
the initial times, both semiclassical and quantum currents look quite
similar. However, already for short times, $\tau =0.32$, the
semicclassical distribution differs from the quantum one. The
semiclassical current only produces a deformation of the
initial distribution, as it follows from the (\ref{Wcl}). The
semiclassically-evolved distribution is slightly narrower than the
quantum one, but still describes very well the effect of spin
squeezing~\cite{Klimov:2002cr}. For longer times, the semiclassical
current keeps twisting the Wigner distribution, which obviously does
not show any sort of interference pattern. The stagnation lines in the
semiclassical case coincide with zeros of the evolved Wigner function,
as it follows from (\ref{cl SL}) and differ from the quantum case,
even at the initial times. Such a difference is significant and can be
in principle used for a detection of genuine quantum features.

The higher moments of the Wigner distribution
\begin{equation}
  \mathfrak{m}_{k} (t) = \left ( \frac{2S+1}{4 \pi} \right )^{k}
  \int_{\mathcal{S}_{2}} d \Omega \, W_{\varrho}^{k} (\Omega |-t)
  \, .
\end{equation}
Since in the semiclassical approximation the evolution is generated by
canonical transformations, these higher moments are time-independent.
The devia- tions from their initial values describe a spread of the
initial distribution due to purely quantum effects. Actually, since
$\partial_{t} \mathfrak{m}(t) |_{t=0} = 0$, the widths of the
$\mathfrak{m}_{k}(t)$ at $t = 0$, given by
$\partial_{t}^{2} \mathfrak{m}(t) |_{t=0} $ define the timescales over
which the semiclassical approximation gives a \emph{bona fide}
description of the dynamics.

\section{Concluding remarks}

In summary, we have studied the dynamics of the Wigner function for
spinlike systems and the associated phase-space flow. For linear
Hamiltonians, the quantum and classical flows coincide. For nonlinear
evolution, there are significant differences between quantum and
classical flows, even for short times $\tau \sim S^{-1/2}$.  From an
experimental viewpoint, quantum effects can hardly be observed by
measuring low-order moments of spin operators~ \cite{Valtierra:2017aa}
when $S\gg 1$. Thus, by analyzing the Wigner current, in
principle, it is possible to detect genuine quantum features of large
spin systems arising in the course of nonlinear dynamics.

\ack
This work is partially supported by the Grant 254127 of CONACyT
(Mexico).  S.~T.~W. is supported by the Ministry of Science and
  Technology Taiwan (Grant MOST 107-2112-M-194-002.
L.~L.~S.~S. acknowledges the support of the Spanish MINECO (Grant
FIS2015-67963-P).

\newpage


\begin{thebibliography}{10}
\expandafter\ifx\csname url\endcsname\relax
  \def\url#1{{\tt #1}}\fi
\expandafter\ifx\csname urlprefix\endcsname\relax\def\urlprefix{URL }\fi
\providecommand{\eprint}[2][]{\url{#2}}

\bibitem{Arnold:1989aa}
Arnold V~I 1989 {\em Mathematical Methods of Classical Mechanics\/} (Berlin:
  Springer)

\bibitem{Berry:1978aa}
Berry M~V 1978 Regular and irregular motion {\em Topics in Nonlinear
  Dynamics\/} ({\em AIP Conf. Proc.\/} vol~46) ed Jorna S p~16

\bibitem{Stratonovich:1956qc}
Stratonovich R~L 1956 {\em JETP\/} {\bf 31} 1012---1020

\bibitem{Berezin:1975mw}
Berezin F~A 1975 {\em Commun. Math. Phys.\/} {\bf 40} 153--174

\bibitem{Hillery:1984aa}
Hillery M, O'Connell R~F, Scully M~O and Wigner E~P 1984 {\em Phys. Rep.\/}
  {\bf 106} 121--167

\bibitem{Lee:1995aa}
Lee H~W 1995 {\em Phys. Rep.\/} {\bf 259} 147--211

\bibitem{Schroek:1996aa}
Schroek F~E 1996 {\em Quantum Mechanics on Phase Space\/} (Dordrecht: Kluwer)

\bibitem{Schleich:2001aa}
Schleich W~P 2001 {\em Quantum Optics in Phase Space\/} (Berlin: Wiley-VCH)

\bibitem{QMPS:2005aa}
Zachos C~K, Fairlie D~B and Curtright T~L (eds) 2005 {\em Quantum mechanics in
  phase space\/} (Singapore: World Scientific)

\bibitem{Moyal:1949aa}
Moyal J~E 1949 {\em Proc. Camb. Phil. Soc.\/} {\bf 45} 99--124

\bibitem{Klimov:2017aa}
Klimov A~B, Romero J~L and de~Guise H 2017 {\em J. Phys. A: Math. Theor.\/}
  {\bf 50} 323001

\bibitem{Ozorio:1998aa}
Ozorio~de Almeida A~M 1998 {\em Phys. Rep.\/} {\bf 295} 265--342

\bibitem{Klimov:2002cr}
Klimov A~B 2002 {\em J. Math. Phys.\/} {\bf 43} 2202--2213

\bibitem{Dittrich:2010aa}
Dittrich T, G{\'o}mez E~A and Pach{\'o}n L~A 2010 {\em J. Chem. Phys.\/} {\bf
  132} 214102

\bibitem{Polkovnikov:2010aa}
Polkovnikov A 2010 {\em Ann. Phys.\/} {\bf 325} 1790--1852

\bibitem{Bauke:2011aa}
Bauke H and Itzhak N~R 2011 {\em arXiv:1101.2683v1\/}

\bibitem{Steuernagel:2013aa}
Steuernagel O, Kakofengitis D and Ritter G 2013 {\em Phys. Rev. Lett.\/} {\bf
  110} 030401

\bibitem{Kakofengitis:2017aa}
Kakofengitis D, Oliva M and Steuernagel O 2017 {\em Phys. Rev. A\/} {\bf 95}
  022127

\bibitem{Oliva:2018aa}
Oliva M, Kakofengitis D and Steuernagel O 2018 {\em Physica A\/} {\bf 502}
  201--210

\bibitem{Donoso:2001aa}
Donoso A and Martens C~C 2001 {\em Phys. Rev. Lett.\/} {\bf 87} 223202

\bibitem{Hughes:2007aa}
Hughes K~H, Parry S~M, Parlant G and Burghardt I 2007 {\em J. Phys. Chem. A\/}
  {\bf 111} 10269--10283

\bibitem{Kakofengitis:2017ab}
Kakofengitis D and Steuernagel O 2017 {\em Eur. Phys. J. Plus\/} {\bf 132}

\bibitem{Friedman:2017aa}
Friedman O~D and Blencowe M~P 2017 {\em arxiv:1703.04844\/}

\bibitem{Skodje:1989aa}
Skodje R~T, Rohrs H~W and VanBuskirk J 1989 {\em Phys. Rev. A\/} {\bf 40}
  2894--2916

\bibitem{Veronez:2013aa}
Veronez M and de~Aguiar M~A~M 2013 {\em J. Phys. A: Math. Theor.\/} {\bf 46}
  485304

\bibitem{Veronez:2016aa}
Veronez M and de~Aguiar M~A~M 2016 {\em J. Phys. A: Math. Theor.\/} {\bf 49}
  065301

\bibitem{Chaturvedi:2006vn}
Chaturvedi S, Marmo G and Mukunda N 2006 {\em Rev. Math. Phys.\/} {\bf 18}
  887--912

\bibitem{Agarwal:1981bd}
Agarwal G~S 1981 {\em Phys. Rev. A\/} {\bf 24} 2889--2896

\bibitem{Brif:1998if}
Brif C and Mann A 1998 {\em J. Phys. A\/} {\bf 31} L9--L17

\bibitem{Heiss:2000kc}
Heiss S and Weigert S 2000 {\em Phys. Rev. A\/} {\bf 63} 012105

\bibitem{Klimov:2000zv}
Klimov A~B and Chumakov S~M 2000 {\em J. Opt. Soc. Am. A\/} {\bf 17} 2315--2318

\bibitem{Klimov:2008yb}
Klimov A~B and Romero J~L 2008 {\em J. Phys. A\/} {\bf 41} 055303

\bibitem{Dowling:1994sw}
Dowling J~P, Agarwal G~S and Schleich W~P 1994 {\em Phys. Rev. A\/} {\bf 49}
  4101--4109

\bibitem{Atakishiyev:1998pr}
Atakishiyev N~M, Chumakov S~M and Wolf K~B 1998 {\em J. Math. Phys.\/} {\bf 39}
  6247--6261

\bibitem{Chumakov:1999sj}
Chumakov S~M, Frank A and Wolf K~B 1999 {\em Phys. Rev. A\/} {\bf 60}
  1817--1822

\bibitem{Chumakov:2000le}
Chumakov S~M, Klimov A~B and Wolf K~B 2000 {\em Phys. Rev. A\/} {\bf 61} 034101

\bibitem{Perelomov:1986ly}
Perelomov A 1986 {\em Generalized Coherent States and their Applications\/}
  (Berlin: Springer)

\bibitem{Varilly:1989ud}
Varilly J~C and Gracia-Bond{\'{\i}}a J~M 1989 {\em Ann. Phys.\/} {\bf 190}
  107--148

\bibitem{Fano:1959ly}
Fano U and Racah G 1959) {\em Irreducible Tensorial Sets\/} (New York: Academic
  Press)

\bibitem{Blum:1981rb}
Blum K 1981 {\em Density Matrix Theory and Applications\/} (New York: Plenum)

\bibitem{Varshalovich:1988ct}
Varshalovich D~A, Moskalev A~N and Khersonskii V~K 1988 {\em Quantum Theory of
  Angular Momentum\/} (Singapore: World Scientific)

\bibitem{Muller:2012ys}
M{\"u}ller C~R, Stoklasa B, Peuntinger C, Gabriel C, {\v R}eh{\'a}{\v c}ek J,
  Hradil Z, Klimov A~B, Leuchs G, Marquardt C and S{\'a}nchez-Soto L~L 2012
  {\em New J. Phys.\/} {\bf 14} 085002

\bibitem{Muller:2016aa}
M{\"u}ller C~R, Madsen L~S, Klimov A~B, S{\'a}nchez-Soto L~L, Leuchs G,
  Marquardt C and Andersen U~L 2016 {\em Phys. Rev. A\/} {\bf 93} 033816

\bibitem{Tilma:2016aa}
Tilma T, Everitt M~J, Samson J~H, Munro W~J and Nemoto K 2016 {\em Phys. Rev.
  Lett.\/} {\bf 117} 180401

\bibitem{Bayen:1978aa}
Bayen F, Flato M, Fronsdal C, Lichnerowicz A and Sternheimer D 1978 {\em Ann.
  Phys.\/} {\bf 111} 61--110

\bibitem{Kitagawa:1993aa}
Kitagawa M and Ueda M 1993 {\em Phys. Rev. A\/} {\bf 47} 5138--5143

\bibitem{Agarwal:1997aa}
Agarwal G~S, Puri R~R and Singh R~P 1997 {\em Phys. Rev. A\/} {\bf 56}
  2249--2254

\bibitem{Corney:2008aa}
Corney J~F, Heersink J, Dong R, Josse V, Drummond P~D, Leuchs G and Andersen
  U~L 2008 {\em Phys. Rev. A\/} {\bf 78} 023831

\bibitem{Rigas:2013aa}
Rigas I, Klimov A~B, S{\'a}nchez-Soto L~L and Leuchs G 2013 {\em New J.
  Phys.\/} {\bf 15} 043038

\bibitem{Oliva:2018ab}
Oliva M and Steuernagel O 2018 {\em arXiv:1811.02952\/}

\bibitem{Valtierra:2017aa}
Valtierra I~F, Romero J~L and Klimov A~B 2017 {\em Ann. Phys.\/} {\bf 383}
  620--634

\end{thebibliography}
\providecommand{\newblock}{}

\end{document}